\documentclass[12pt]{amsart}
\newtheorem{theo}{Theorem}
\newtheorem{cor}[theo]{Corollary}
\newtheorem{lem}[theo]{Lemma}
\newenvironment{rem}{\vspace{.25cm}\addtocounter{theo}{1} 
\noindent {\bf Remark {\thetheo}}.}{\vspace{.25cm}} 
\newenvironment{defin}{\vspace{.25cm}\addtocounter{theo}{1} 
\noindent {\bf Definition {\thetheo}}.}{\vspace{.25cm}}
\newenvironment{pr}{\vspace{.25cm} \noindent \textit{Proof}}{\hfill 
{\raggedleft 
} \vspace{.25cm}} 

\newenvironment{exam}{\vspace{.25cm}\addtocounter{theo}{1} 
\noindent {\bf Example {\thetheo}}.}{\vspace{.25cm}}

\begin{document}

\centerline{Journal of Integer Sequences, \hfill SPhT/99/134}
\centerline{Vol 3, (2000), Article 00.1.4 \hfill cond-mat/0003049}

\vskip 2cm
\centerline{\LARGE\bf On the kernel of tree incidence matrices}
\vskip 1.5cm

\centerline{\large M. Bauer and O. Golinelli}

\bigskip 
\centerline{Service de Physique Th\'eorique, CEA Saclay,} 
\centerline{F-91191, Gif-sur-Yvette, France.}
\medskip
\centerline{email : bauer and golinelli @spht.saclay.cea.fr}

\vskip2.5cm
\centerline{\bf{Abstract}}
\textit{
We study the height of the delta peak at 0 in the spectrum of incidence
matrices of random trees. We show that the average fraction of the spectrum
occupied by the eigenvalue $0$ in a large random tree is asymptotic to
$2x_*-1=0.1342865808195677459999\cdots$ where $x_*$ is the unique real root
of $x = e^{-x}$.  For finite trees, we give a closed form, a generating
function, and an asymptotic estimate for the sequence $(z_n)_{ n \geq
1}=1,0,3,8,135,1164,21035\cdots$ of the total multiplicity of the
eigenvalue 0 in the set of $n^{n-2}$ tree incidence matrices of size $n$.
} 

\section{Introduction.}

By a classical result in graph theory, the number of labeled
trees\footnote{Precise definitions for this and the following terms
can be found in Section \ref{sec:defs}.} on 
$n \geq 1$ vertices is $n^{n-2}$. We endow the set ${\mathcal T}_n$ of
labeled trees
on $n\geq 1$ vertices with the uniform probability, giving weight
$n^{2-n}$ to each tree. 

Each tree in ${\mathcal T}_n$ comes with its incidence matrix,
the $n \times n$ matrix with entry $ij$ equal to $1$ if there is an
edge between vertices $i$ and $j$ and to $0$ else. Each such
(symmetric) matrix has $n$ (real) eigenvalues, which by definition
form the spectrum of the corresponding tree. This leads in turn to
$n\, n^{n-2} =n^{n-1}$ eigenvalues counted with multiplicities for
${\mathcal T}_n$ as a whole. In the sequel, we wish concentrate on
the multiplicity of the eigenvalue $0$. Let $Z(T)$ be the
multiplicity of the eigenvalue $0$ in the spectrum of the incidence
matrix of the tree $T$, i.e. the dimension of the kernel. For each
$n\geq 1$, the restriction $Z_n$ of 
$Z$ to ${\mathcal T}_n$ is a random variable. We set
$z_n=\sum_{T \in {\mathcal T}_n} Z_n(T)$. The expectation of $Z_n(T)$ is
$\mathbb{E}(Z_n) =z_n/n^{n-2}$.

To illustrate these definitions, we give a direct counting of
$z_1,\cdots,z_4$ in appendix \ref{app:exam}.

\vspace{.3cm}

Our aim is to prove :

\begin{theo} \label{theo} Let $z_n$ be the total multiplicity of the
eigenvalue $0$ in the spectra of the $n^{n-2}$ labeled trees on $n$
vertices. Then~:

i) Closed form :
\begin{eqnarray*}z_n & = & n^{n-1}-2\sum_{2 \leq m \leq n}(-1)^m
n^{n-m}m^{m-2}\binom{n-1}{m-1} \\ 
 \frac{z_n}{n^{n-2}} & \equiv & \mathbb{E}(Z_n)  \; =  \; n\left(1-2\sum_{2
\leq m \leq n}\frac{(-1)^m}{m} 
\left(\frac{m}{n}\right)^m \binom{n}{m}\right).
\end{eqnarray*}

ii) Formal power series identity :
$$ x^2+2x-xe^x=\sum_{n \geq 1} \frac{z_n}{n!}\left(xe^x e^{-xe^x}\right)^n.$$
\end{theo}

and 

\begin{cor} \label{cor} For large $n$, $\mathbb{E}(Z_n)$ has an
asymptotic expansion 
in powers of $1/n$, whose first two terms are
$$\mathbb{E}(Z_n) = (2x_*-1)n + \frac{x_*^2(x_*+2)}{(x_*+1)^3} +O(1/n)$$ 
where $x_*=0.5671432904097838729999\cdots$ is the unique real root of 
$x=e^{-x}$. In
particular, the average fraction of the spectrum occupied by the
eigenvalue $0$ in a large random tree is asymptotic to
$2x_*-1=0.1342865808195677459999\cdots$.  
\end{cor}

\begin{rem} We do not try to justify here that fluctuations in random
trees become small when the number of vertices is large. However, it
is expected that $\mathbb{E}(Z_n^2)-\mathbb{E}(Z_n)^2$ grows only linearly with
the number of vertices, so that in an appropriate sense the 
fraction of the spectrum occupied by the eigenvalue $0$ in an infinite 
random tree is $2x_*-1$ with probability 1. 
\end{rem}

\begin{rem} With the explicit formula above, it is easy to list the
first terms in the sequence $(z_n)_{n\geq 1}$, which are
$$1,0,3,8,135,1164,21035,322832,7040943,153153620,4048737099,\cdots$$
\end{rem}

To prove part \textit{i)} of Theorem \ref{theo} we establish a few
preparatory lemmas of independent interest. Then we prove \textit{ii)}
using Lagrange inversion and get Corollary \ref{cor} with
the steepest descent method.

But first, we need to fix conventions and notations.

\section{Definitions.} \label{sec:defs}

Even if we are interested ultimately only  
in trees, we shall need more general graphs (for instance, forests) in 
the proofs, so we give for the sake of completeness a collection
definitions. Most of them are standard, and the reader is encouraged
to skip this this section and come
back to it only when needed. The fundamental definition is

\begin{defin} 
A \textit{simple graph} $G$ is a pair
$(V,E)$ where $V$ is a finite set called the set of \textit{vertices} 
and $E$ is a subset of $V^{(2)}\equiv\{\, \{x,y\}, \,x \in V,\, y \in V,
\,x\neq y\}$ called the set of \textit{edges}.
\end{defin}

\begin{rem} The adjective \textit{simple} refers to the fact that
there is at most \textit{one} edge between two vertices and that edges are
pairs of \textit{distinct} vertices. As we have no use of more general 
graphs in the sequel, we shall from now on use \textit{graph} 
for \textit{simple graph}.
\end{rem}

\begin{defin} If $V$ is empty, then we say that the graph $G$ is
\textit{empty}. If $\{x,y\}$ belongs to $E$, we say that there is an
edge between $x$ and $y$ and that
$x$ and $y$ are \textit{adjacent vertices} in $G$. The vertices adjacent
to a given vertex $x$ are called the \textit{neighbors} of $x$. The
number of neighbors of a vertex $x$ is called the \textit{degree} of
$x$. A \textit{leaf} of $G$ is a vertex of degree $1$. Two edges of
$G$ with a common vertex are called \textit{adjacent edges} 
\end{defin} 

\begin{defin} A \textit{labeled} graph on $n \geq 1$ vertices is a
graph with vertex set $[n]=\{1,\cdots,n\}$.
The \textit{incidence matrix} of a labeled graph on $n$ vertices 
is the $n \times n$ matrix with entry $ij$ equal to $1$ if 
there is an edge between vertices $i$ and $j$ and to $0$ else. 
\end{defin}

\begin{rem} If the graph $G$ has $|V|=n \geq 1$ vertices\footnote{For
any finite 
set $S$ , $|S|$ is the number of elements in $S$.}, any bijection
between $V$ and $[n]$ defines a labeled graph. The incidence matrices 
for different bijections differ only by a permutation of the lines and 
columns. In particular the eigenvalues are independent of the
bijection. They are real because, by construction, incidence matrices
are symmetric.
\end{rem}

\begin{defin} The \textit{spectrum} of a graph is the set of eigenvalues 
(counted with multiplicities) of any of the associated incidence
matrices. By convention, the spectrum of the empty graph is empty. 
\end{defin}

\begin{defin}
A \textit{subgraph} of a graph $G=(V,E)$ is a graph $(W,F)$ such that
$W \subset V$ and $F \subset E $. An \textit{induced subgraph}
of $G$ is a graph $(W,F)$ such that $W \subset V$ and $F = E \cap W^{(2)}$. 
\end{defin}

\begin{defin}
We say that two vertices $x$ and $x'$ $\in V$ are in the \textit{same
component} of $G$ if there is a sequence $x=x_1,\cdots,x_n=x'$ in
$V$ such that adjacent terms in the sequence are adjacent in
$G$ (taking $n=1$ shows that luckily $x$ and $x$ are in the same
component). This gives a partition of $V$. Each component defines an
induced subgraph of $G$ which is called a \textit{connected component} of
$G$. Then $G$ can be thought of as the disjoint union of its connected 
components. We say that $G$ is \textit{connected} if it has only one
connected component.  
\end{defin}

\begin{defin} A \textit{polygon} in a graph $G$ is a sequence
$x_0,x_1,\cdots,x_n$, $n \geq 3$ of vertices such that adjacent terms
in the sequence are adjacent in $G$, $x_0=x_n$ and $x_1,\cdots,x_n$
are distinct.
\end{defin}

\begin{defin} A \textit{forest} is a graph without polygons. A
\textit{tree} is a non-empty connected forest.
\end{defin}

\begin{rem} Clearly a subgraph of a forest is a forest. The connected
components of a non empty forest are trees. One shows
easily that that  
a tree with $n \geq 2$ vertices has at least two leaves. Then a simple
induction shows that a tree is exactly a connected graph for which the 
number of vertices is $1$ plus the number of edges. A classical
theorem of Cayley states that there are $n^{n-2}$ labeled trees on $n$ 
vertices(see for instance proposition 5.3.2 in \cite{stanley}).
\end{rem}

\section{Two preparatory lemmas.}

The first lemma is a characterization of the dimension of the kernel
of incidence matrices viewed as a function on forests.

\begin{lem} \label{lem:zero} The function $Z$ which associates to any forest
the multiplicity of the eigenvalue $0$ in its spectrum is
characterized by the following properties :

i) The function $Z$ takes the value $0$ on $\emptyset$, the empty forest.

ii) The function $Z$ takes the value $1$ on \begin{picture}(10,4)
\put(5,2){\circle*{2}} \end{picture}, the forest with one vertex.

iii) The function $Z$ is additive on disjoint components, i.e. if the
forest $F$ is the union of two disjoint forests $F_1$ and $F_2$ then
$Z(F)=Z(F_1)+Z(F_2)$

iv) The function $Z$ is invariant under ``leaf removal'', i.e. if $x$
is a leaf of $F$, $y$ its (unique) $neighbor$, $V'=V \backslash \{x,y\}$,
and $F'$ is the subforest of $F$ induced by $V'$ then $Z(F)=Z(F')$.
\end{lem}

\begin{rem} That the function $Z$ fulfills properties \textit{i)--iv)}
was no doubt known decades ago (see for instance section 8.1, 
H\"uckels theory, in \cite{cvetkovic}). We give a proof, because in
the sequel we want to emphasize and use the simple fact that these properties
characterize the function $Z$.
\end{rem}
 
\begin{pr} \textit{of Lemma \ref{lem:zero}}. First, we show that the function
$Z$ has properties \textit{i)--iv)}. In fact, this is true for general
graphs (not only forests). Properties \textit{i)} and \textit{ii)}
follow from the definition of $Z$, property \textit{iii)} follows from 
the fact that the incidence matrix can be put in block diagonal form,
each block corresponding to a connected component. Property
\textit{iv)} is only slightly more complicated. With an 
appropriate labeling of the vertices, the incidence matrix
$\mathbf{M}$ of $F$ can be decomposed as 
$$\mathbf{M}=\left(\begin{array}{ccc} 0 & 1 & \mathbf{0} \\
1 & 0 & \mathbf{N} \\
\mathbf{0} & ^t\mathbf{N} & \mathbf{M'} \end{array}\right)$$
where the first line and column are indexed by the leaf $x$, the second 
line and column is indexed by its neighbor $y$, $\mathbf{N}$ describes the
edges between this neighbor and $V'$, and $\mathbf{M'}$ is the
incidence matrix for $V'$. Then $\mathbf{v}= \, ^t \! (v_1,v_2,\mathbf{v'})$ is
in the kernel of $\mathbf{M}$ if and only if
\begin{eqnarray*} v_2 & = & 0 \\
v_1 & =& - \mathbf{Nv'} \\
\mathbf{M'v'}& = & -^t\mathbf{N}v_2.
\end{eqnarray*}    
So $v_2=0$ which reported in the third equation gives
$\mathbf{M'v'}=\mathbf{0}$ implying that $\mathbf{v'}$ is in the kernel of
$\mathbf{M'}$, and then the second equation just tunes $v_1$ the
appropriate value. So the kernels of $\mathbf{M}$ and $\mathbf{M'}$
have the same dimension. This proves \textit{iv)}.

Now, any tree with more than $1$ vertex has leaves, so leaf removal
as defined in \textit{iv)} allows to reduce the forest $F$ to a
(possibly empty) family of isolated vertices (all connected
components have only one vertex). Hence, there is at most one function, 
namely $Z$, that can satisfy properties \textit{i)--iv)}. 
\end{pr}

\begin{rem} Leaf removal and additivity give an efficient algorithm to 
compute the multiplicity of the eigenvalue 0 for a given forest,
especially when this forest is given as a drawing. 
\end{rem}

The next lemma gives a practically awful but theoretically useful
formula for the function $Z$.

\begin{lem} \label{lem:edges} Let $L$ be the function on forests defined by 

i') The function $L$ takes value $0$ on $\emptyset$, the empty
forest. 

ii') The function $L$ takes value $1$ on $\begin{picture}(10,4)
\put(5,2){\circle*{2}} \end{picture}$, the forest with one
vertex.

iii') The function $L$ takes value $0$ on disconnected forests.

iv') The function $L$ takes value $2(-1)^{n-1}$ on trees with $n \geq
2$ vertices.

Then, for any forest $F$
$$Z(F)=\sum_{F' \subset F} L(F')= \sum_{T' \subset F} L(T')$$
where the first sum is over induced subforests of $F$, and the second
over induced subtrees of $F$.
\end{lem}

\begin{rem} \label{rem:non-adj} For a given forest, there is a much
nicer formula,
directly connected to the geometry of the forest (again, see for
instance section 8.1, H\"uckels theory, in \cite{cvetkovic}). In fact,
let $Q(F)$ be the maximum among the cardinals of sets of pairwise non-adjacent
edges in $F$, and $N(F)$ be the number of vertices in $F$.
Then $Z(F)=N(F)-2Q(F)$. It is easy to show that
 $N(F)-2Q(F)$ satisfies properties \textit{i)--iv)} of Lemma
 \ref{lem:zero}. In particular, a possible way to maximize the number
 of non-adjacent edges in $F$ in the situation \textit{iv)} is to do so 
 on $F'$ and add the edge $\{x,y\}$. Anyway, this explicit formula allows
us to restate our 
theorems in terms of the random variable $Q_n$, the restriction of $Q$ to
${\mathcal T}_n$. For instance, in a large random tree on $n$
vertices, one can find about $(1-x_*)n$ pairwise non-adjacent edges. Note that
$1-x_*=0.4328567095902161270000\cdots$ is not much smaller than $0.5$ (the
upper bound for $Q(T)/N(T)$ for a given tree because $Z(T)=N(T)-2Q(T)$
is always nonnegative).  
\end{rem}

\begin{pr} \textit{of Lemma \ref{lem:edges}}. Our strategy is to use the
characterization of $Z$ in 
Lemma \ref{lem:zero}. First, we observe that the second equality is
a trivial consequence of \textit{i')} and \textit{iii')}.
We define a new function $Z'$ on the set of forests by
$$Z'(F)\equiv \sum_{T' \subset F} L(T')$$
(where the sum is over induced subtrees of $F$) and show that $Z'$
satisfies properties \textit{i)--iv)} of Lemma \ref{lem:zero}. 

As the empty forest has no non-empty induced
subtree \textit{i')} implies \textit{i)}. 

In the same vein, the forest with one vertex as only one non-empty
induced subtree, namely itself, so \textit{ii')} implies \textit{ii)}. 

If the forest $F$ is the union of two disjoint forests
$F_1$ and $F_2$, an induced subtree of $F$ is either an induced
subtree of $F_1$ or an induced subtree of $F_2$, and the sum defining
$Z'(F)$ splits as $Z'(F_1)+Z'(F_2)$, showing that $Z'$ satisfies
property \textit{iii)}.

Now, if $x$ is a leaf of $F$ and $y$ its neighbor, we define $V'=V
\backslash \{x,y\}$, $V''=\{x,y\}$ and consider 
$F'$ and $F''$, the subforests of $F$ induced by $V'$ and $V''$
respectively. We split the sum
defining $Z'(F)$ in three pieces. The first is over the induced subtrees
of $F'$. This is just the sum defining $Z'(F')$. The second is over
the induced subtrees of $F''$, which is a tree on two vertices. Its
subtrees are itself, with weight $L(F'')=2(-1)^{2-1}=-2$, and two
trees with one vertex, each with weight $L(\begin{picture}(4,4)
\put(2,2){\circle*{2}} \end{picture})=1$, so this second sum
gives $0$. The third sum is over induced subtrees that have vertices in both
$V'$ and $V''$. If this sum is not empty, every tree that appears in
it has $y$ as a vertex (by connectivity) and has at least two vertices
(because the tree consisting of $y$ alone has already been
counted). Then we can group these trees in pairs, a tree containing
$x$ being paired with the same tree but with $x$ and the edge $\{x,y\}$
deleted. The function $L$ takes opposite values on the two members of
a pair, so the third sum contributes $0$. Hence $Z'$ satisfies
property \textit{iV)}. So $Z'(F)=Z'(F')$.
\end{pr}

\begin{rem} These two lemmas have an obvious extension to bicolored
forests. If we use black and white as colors, and count the zero
eigenvectors having value zero on white vertices, we only need to
replace \textit{ii)} in Lemma 
\ref{lem:zero} by

\textit{ii) The function $Z$ takes value $1$ on $\bullet$, the forest
with one vertex colored in black and $0$ on $\circ$, the forest with
 one vertex colored in white.} 

\noindent and \textit{ii')} and \textit{iii')} in Lemma \ref{lem:edges} by

\textit{ii')  The function $L$ takes value $1$ on $\bullet$, the forest
  with one vertex colored in black and $0$ on $\circ$, the forest with
  one vertex colored in white.}

\textit {iii') The function $l$ takes value $(-1)^{n-1}$ on trees with
$n \geq 2$ vertices.} 

The proofs remain the same.
\end{rem}

\begin{rem} \label{rem:inclexcl} The formula
$$Z(F)=\sum_{F' \subset F} L(F')$$ 
can be inverted using inclusion-exclusion to give 
$$L(F)=\sum_{F' \subset F}(-1)^{|V(F)|-|V(F')|} Z(F').$$ 
This identity has an application in random graph theory \cite{nous},
which is why we got interested in Lemma \ref{lem:edges} in the first
place. 
\end{rem}

\section{Main proofs.}

We have now the necessary tools to prove theorem \ref{theo}. 

\begin{pr} \textit{of Theorem \ref{theo}}. By Lemma \ref{lem:zero} 
$$z_n \equiv \sum_{T \in {\mathcal T}_n} Z_n(T) = \sum_{m=1}^n \sum_{T
\in {\mathcal T}_n} \sum_{T' \in {\mathcal T}_m}^{T' \subset T} L(T').$$
As the
function $L$ depends only on the number of vertices, for fixed $m$ the 
double sum $\sum_{T \in {\mathcal T}_n} \sum_{T' \in {\mathcal
T}_m}^{T' \subset T}$ is simply
a multiplicity. We count this multiplicity as follows : we remove from
$T$ the edges of $T'$, so we are left with m trees, each with a
special vertex, the one belonging to $T'$. This is by definition what
is called a planted forest (or rooted forest) with $n$ vertices and
$m$ trees. The number  
of such objects is $m\binom{n}{m}n^{n-m-1}$
(see for instance proposition 5.3.2 in \cite{stanley}). Conversely,
starting from such a planted forest with $m$ trees (each with a
special vertex) and $n$ vertices, we can build a tree on the special 
vertices in $m^{m-2}$ ways. So
$$\sum_{T \in {\mathcal T}_n} \sum_{T' \in {\mathcal T}_m}^{T' \subset
T} 1 = m^{m-1}\binom{n}{m}n^{n-m-1}.$$

Hence summation over $m$ gives 
$$z_n = n^{n-1}-2\sum_{2 \leq m \leq n}(-1)^m
n^{n-m-1}m^{m-1}\binom{n}{m}.$$
Simple rearrangements lead to the two equivalent formul\ae\ in
\textit{i)}, the first one making clear that $z_n$ is an integer.

To obtain the generating function in \textit{ii)}, we need a mild
extension of the Lagrange inversion formula (see for instance section 5.4 in
\cite{stanley}), which states that if $f(x)$ is a formal power series
in $x$ starting as $f(x)=x+O(x^2)$ and $g(x)$ is an arbitrary formal
power series in $x$, 
$$\left(g\circ f^{-1}\right)(t)=g(0)+\sum_{n\geq
1}\frac{1}{n}\left[\frac{x^ng'(x)}{f(x)^n}\right]_{n-1} t^n.$$
where $[h(v)]_k$ is by definition the $k^{th}$ coefficient of the
formal power series $h(v)$.

As an immediate application, we see that if $t=xe^x$ then
$$x=\sum_{m\geq 1} (-m)^{m-1}\frac{t^m}{m!}$$ and 
$$-x-x^2/2=\sum_{m\geq 1} (-m)^{m-2}\frac{t^m}{m!}.$$ 

Now we introduce $y=te^{-t}$ and define a sequence $z'_n,n\geq1$ by
$$x^2+2x-xe^x=\sum_{n\geq 1} z'_n \frac{y^n}{n!},$$
but instead of applying directly the Lagrange inversion formula to
$y=xe^x e^{-xe^x}$, we first substitute the $t$-expansion (already
obtained by Lagrange inversion) on the left-hand side which yields 
$$-2\sum_{m \geq 1} (-m)^{m-2}\frac{t^m}{m!}-t,$$
and then apply
Lagrange inversion with $y=te^{-t}$. The result is
$$\frac{z'_n}{n!}=\frac{1}{n}\left[e^{nt}\left(1-2\sum_{m\geq
2}\frac{(-m)^{m-2}}{(m-1)!} t^{m-1} \right)\right]_{n-1}.$$
Straightforward expansion of this formula shows that $z'_n=z_n$, and
this proves the generating function representation in
\textit{ii)}. 
\end{pr}

\begin{rem} The derivation of \textit{ii)} is quite artificial. It 
turns out that random graph theory gives a natural proof \cite{nous}
using the formula mentioned in remark 22.
\end{rem}

\begin{pr} of Corollary \ref{cor}. This time we use Lagrange inversion with 
$y=xe^x e^{-xe^x}$, in a contour integral representation\footnote{We
include the factor $\frac{1}{2i\pi}$ in the symbol $\oint$.}. So
$$\frac{z_n}{n!}=\frac{1}{n}\oint \frac{dx}{(xe^x e^{-xe^x})^n}
(1+x)(2-e^x),$$ 
where the contour is a small anticlockwise-oriented circle around the
origin. For large $n$ we use the steepest descent method to obtain the 
asymptotic expansion of $z_n$.
As $\frac{d}{dx}xe^x e^{-xe^x}=(1+x)(1-xe^x)e^x e^{-xe^x}$, the
saddle points of $xe^x e^{-xe^x}$ are $x=-1$ and the solutions to
$x=e^{-x}$. This equation has a unique real
root, $x_*$, which is positive. Numerically,
$x_*=0.5671432904097838729999\cdots.$
On the other hand, $x=e^{-x}$ has an infinite number of complex
solutions, coming in complex conjugate pairs. Asymptotically, the
imaginary parts of these zeroes are evenly spaced by about $2\pi$,
while their real parts are negative and grow logarithmically in
absolute value. Consideration of the
landscape produced by the modulus of the function $xe^x e^{-xe^x}$
shows that the small circle around the origin can be deformed to give
the union of two steepest descent curves, one passing through $x=-1$
and the other through $x=x_*$. These two curves are asymptotic to the
two lines $y=\pm \pi$ at $x \rightarrow +\infty$. Hence, despite the
fact that the value
of $xe^x e^{-xe^x}$ is the same, namely $1/e$, at all the complex saddle
points and at $x_*$, the complex saddle points do not contribute to
the asymptotic expansion of $z_n$ at large $n$. Moreover, the point
$x=-1$ only gives subdominant contributions because
$-e^{-1}e^{e^{-1}}$ is larger than $1/e$ in absolute value. So we
concentrate on the asymptotic expansion around $x_*$.  
As $$\log xe^x e^{-xe^x}= -1-\frac{(x_*+1)}{2x_*}(x-x_*)^2+O((x-x_*)^3)$$
we infer that
$$e^{-n}\sqrt{2\pi n}\oint \frac{dx}{(xe^x e^{-xe^x})^n}
(1+x)(2-e^x)$$
 has an asymptotic expansion in powers of $1/n$. Hence,
by use of Stirling's formula for $n!$, we conclude that 
$\mathbb{E}(Z_n)=z_n/n^{n-2}$ has an asymptotic expansion 
in powers of $1/n$. The first two terms are obtained by brute force.
\end{pr}

\appendix

\section{Examples of direct multiplicity counting.} \label{app:exam}

This appendix gives the counting of trees and
multiplicities of $0$ in the spectrum for trees on $n=1,2,3$ or $4$ vertices. 

\begin{exam} For $n=1$ there is only one tree, \begin{picture}(10,4)
\put(5,2){\circle*{2}} \end{picture},
and one way to label it, giving a total of $1=1^{1-2}$ tree on one vertex.
The incidence matrix is $(0)$, so the eigenvalue $0$ occurs with
multiplicity $z_1=1$.
\end{exam}

\begin{exam} For $n=2$ there is only one tree, \begin{picture}(20,4)
\put(5,2){\line(1,0){10}} \put(5,2){\circle*{2}}
\put(15,2){\circle*{2}} \end{picture}, and one way to label it,
giving a total of $1=2^{2-2}$ tree on two vertices.   
The incidence matrix is
$$\left(\begin{array}{cc} 0 & 1 \\ 1 & 0 \end{array}\right),$$ so the
eigenvalue $0$ occurs with 
multiplicity $z_2=0$.
\end{exam}

\begin{exam} For $n=3$ there is only one tree, \begin{picture}(30,4)
\put(5,2){\line(1,0){20}} \put(5,2){\circle*{2}} \put(15,2){\circle*{2}}
\put(25,2){\circle*{2}} \end{picture}, and three ways to label it,
giving a total of $3=3^{3-2}$ trees on three vertices.
Up to permutation of rows and columns, the incidence matrix  for each
of these three labeled trees is 
$$\left(\begin{array}{ccc} 0 & 1 & 0 \\ 1 & 0 & 1 \\ 0 & 1 & 0
\end{array}\right),$$ which has zero as an eigenvalue with
multiplicity $1$ (a corresponding eigenvector is $^t(1,0,-1)$), so
there is a total of $3\times 1$ zero eigenvalues, and $z_3=3$ 
\end{exam} 

\begin{exam} For $n=4$ there are two trees, \begin{picture}(40,4)
\put(5,2){\line(1,0){30}} \put(5,2){\circle*{2}} \put(15,2){\circle*{2}}
\put(25,2){\circle*{2}} \put(35,2){\circle*{2}}\end{picture} ($12$
ways to label it), and \begin{picture}(30,14) \put(5,2){\line(1,0){20}}
\put(15,2){\line(0,1){10}}\put(5,2){\circle*{2}} \put(15,2){\circle*{2}} 
\put(25,2){\circle*{2}} \put(15,12){\circle*{2}}\end{picture} ($4$
ways to label it), giving a total of $12+4=16=4^{4-2}$ trees on
three vertices. Up to permutation of rows and columns, the two
incidence matrices are
$$\left(\begin{array}{cccc} 0 & 1 & 0 & 0 \\ 1 & 0 & 1 & 0 \\ 0 & 1 &
0 & 1 \\ 0 & 0 & 1 & 0 \end{array}\right) \qquad \mbox{and} \qquad
\left(\begin{array}{cccc} 0 & 1 & 0 & 0 \\ 1 & 0 & 1 & 1 \\ 0 & 1 & 
0 & 0 \\ 0 & 1 & 0 & 0 \end{array}\right).$$
The first does not have $0$ as an eigenvalue, whereas the second has
zero as an eigenvalue with multiplicity $2$ (corresponding 
eigenvectors are for instance $^t(1,0,-1,0)$ and $^t(1,0,0,-1)$), so
there is a total of $12\times 0 + 4 \times 2$ zero eigenvalues, and
$z_4=8$. 
\end{exam}


\begin{thebibliography}{99}
\bibitem{cvetkovic} D.-M. Cvetkovi\'{c}, M. Doob and H. Sachs,
\textit{Spectra of Graphs}, Academic Press, New York, 1980.
\bibitem{nous} M. Bauer and O. Golinelli,  
\textit{On the spectrum of random graphs}, in preparation.
\bibitem{stanley} R.-P. Stanley, 
\textit{Enumerative Combinatorics, Vol II}, Cambridge University
Press, Cambridge, 1997. 
\end{thebibliography}
\end{document}